\documentclass[12pt]{article}
\def\by#1#2{\frac{\displaystyle {#1}}{\displaystyle {#2}}}
\def\d{{\rm d}}
\usepackage{graphicx}
\usepackage{textcomp}
\usepackage{mathcomp}
\usepackage{amsmath}
\usepackage{amssymb}
\begin{document}

\begin{flushright}
December 30, 2008 \\
IMSc/2008/12/18
\end{flushright}

\begin{center}
{\huge {\bf An extended model for nonet pseudo-scalar
meson fragmentation }} \\ [0.3cm]
{\large D. Indumathi, Basudha Misra} \\ [0.3cm]
{\large The Institute of Mathematical Sciences, Chennai, India} \\
\end{center}

\paragraph{Abstract}: An SU(3) symmetric model with high predictivity
for octet meson ($\pi, K$) quark fragmentation functions with a simple
approach to SU(3) symmetry breaking (due to the relatively heavy strange
quarks) is extended to the singlet sector, with some reasonable
assumptions, in order to study $\eta$ and $\eta'$ fragmentation.
Due to the presence of SU(3) symmetry, fits to the
$\pi$ and $K$ data help to constrain the fragmentation functions
for $\eta$ and $\eta'$ mesons. The use of 2-jet and 3-jet (especially the
gluon jet) inclusive meson production in $e^+\,e^-$ collisions and
$\pi$, $\eta$ inclusive production in $p\,p$ collisions enable the
extraction of the gluon fragmentation functions as well. While sea
quarks in strange mesons ($K^\pm$, $K^0$, $\overline{K}^0$) and the
heavier $\eta$, $\eta'$ mesons are suppressed by a factor of $\lambda_H =
m_\pi^2/m_H^2 \lesssim 0.1$ for $H=K, \eta, \eta'$, the gluons are not as
severely suppressed: $f_g^H \sim 0.3$--0.35. A detailed parametrisation
of the three independent fragmentation functions $V(x, Q^2)$, $\gamma(x,
Q^2)$, and $D_g(x, Q^2)$ are given in LO QCD; {\em all nonet meson gluon
fragmentation functions and quark fragmentation functions of all flavours}
can be expressed in terms of these functions with a few model parameters
including $\lambda_H$ and $f_g$. The data prefer a nonet mixing angle
$-24^\circ \le \theta_P \le -16^\circ$ and rule out no-mixing.

\vspace{0.2cm}


\section{Introduction} 

The PHENIX experiment at RHIC \cite{phenixeta} has measured
inclusive transverse momentum spectra of $\eta$ mesons in the range
$p_T \approx 2$--11 GeV at mid-rapidity $(|y|\sim|\eta| < 0.35 )$ in
$p\, p$, $d\, Au$ and $Au\, Au$ collisions at $\sqrt{s_{NN}}=200$ GeV.
Combining with earlier data on $\pi$ production \cite{phenixpi}, as
well as other similar data \cite{otherdata}, their analysis shows that
the $\eta/\pi^0$ production ratio is roughly constant, $\approx 0.5 $,
for a range of center-of-mass energies ($ \sqrt{s_{NN}} \approx 3$--1800
GeV). This number agrees with $e^+e^-$ annihilation data at $ \sqrt{s}
= 91.2 $ GeV for high scaled momentum $x_p$. This ratio is interesting
because of its potential as a good signal for quark-gluon-plasma
(QGP). Hence, a good base-line study of $\eta$ fragmentation functions
in $p\,p$ collisions is essential for use as a normalisation factor in
the study of this signal of QGP.

There are many previous studies on meson fragmentation functions at both
leading order (LO) and next-to-leading order (NLO) in QCD. While the
fits at NLO on $\pi$ and $K$ inclusive production presented in
Ref.~\cite{KKP} are classic and detailed, several updated approaches
have been studied \cite{otherpi}. While these have concentrated on the
dominantly produced $\pi$ and $K$ mesons which constitute the bulk of the
particle content of jets, for example, about 91\% at the $Z$-pole,
$\eta$ (and $\eta'$) fragmentation has not been much studied, as has been
pointed out in Ref.~\cite{phenixeta}: one such study is
Ref.~\cite{othereta}.

In our earlier work \cite{IMR}, an SU(3) model of light quark $(u,d,s)$
fragmentation functions for octet baryons and pseudo-scalar mesons
which is in good agreement with hadro production ($\pi$ and $K$) data
in $e^+e^-$ annihilation experiments was developed. Due to the presence
of SU(3) symmetry, the model can predict the (purely octet part of the)
$\eta$ fragmentation functions; in fact, the pure octet $\eta_8$
fragmentation functions were listed in this paper. However, due to
octet--singlet mixing, the $\eta$ and $\eta'$ mesons were not considered
in this paper. The RHIC/PHENIX result motivates us to extend this model
to the singlet sector of the pseudo-scalar mesons.

One of the main advantages of our approach is the presence of SU(3)
symmetry so that the valence and sea quark fragmentation functions of
the octet mesons are related to each other and can be written in terms
of just three independent fragmentation functions, a valence function
$V(x, Q^2)$, a sea function $\gamma(x, Q^2)$ and the gluon $D_g(x,
Q^2)$. The model thus has high predictive power. SU(3) breaking effects
and singlet--octet mixing are then included by introducing several
parameters, which are taken to be constants in this simple-minded approach
and yet give good agreement with data. In particular, it is possible
to greatly improve the quality of the fits by using the abundant data
on $\pi$ and $K$ inclusive production to fix the quark and gluon
fragmentation functions and so to obtain both the gluon fragmentation
function and quark fragmentation functions for all flavours for the
$\eta$ meson.

In the next section, we briefly review the cross-sections and rates for
inclusive particle production in $e^+\,e^-$ (2-jet and 3-jet) and $p\,p$
collisions. In section 3, we highlight the salient features of the model, which is based on SU(3)
flavour symmetry of the light quarks, $u$, $d$, and $s$. A singlet
SU(3) symmetry breaking parameter, $\lambda \sim m_\pi^2/m_K^2$,
incorporates the effects of strangeness suppression in the model,
which has good predictivity for all the octet $\pi$ and $K$ meson
fragmentation functions. The model is then extended to include the
SU(3) singlet meson, along with additional parameters $\theta_P$, $f_d$,
$f_g$, that describe the singlet--octet mixing. In Section 4, the detailed
phenomenology is carried out, by fitting the free parameters in comparison
with known data in both the $e^+\,e^-$ and $p\,p$ sectors. Section
5 contains a detailed discussion and a summary of the conclusions.

\section{Cross sections and rates}
We consider the production of pseudo-scalar mesons in $e^+\,e^-$
annihilation and $p\,p$ scattering processes to leading order in
perturbative QCD. We focus primarily on the production of ($\pi$, $K$),
$\eta$ and $\eta'$ mesons.
\nopagebreak

\subsection{$e^+\,e^-$ processes}
To leading order, the hadro-production cross-section in the $e^+\,e^-$
scattering process at c.m. energy $\sqrt{s}$ is \cite{pdg}:
\begin{eqnarray}
\by{1}{\sigma_{tot}}\by{\d \sigma^h}{\d x} & = & \by{\sum_q c_q
D_q^h(x, Q^2)}{\sum_q c_q}~.
\label{eq:totcross}
\end{eqnarray}
Here $c_q$ are the charge factors associated with a quark $q_i$ of flavour
$i$ and can be expressed \cite{pdg} in terms of the electromagnetic
charge, $e_i$, and the vector and axial vector electroweak couplings,
$v_i = T_{3i}-2 e_i \sin^2 \theta_{\rm w}$ and $a_i = T_{3i}$, as
\begin{eqnarray} \nonumber
c_q & = & c_q^V + c_q^A~, \\ \nonumber
c_q^V & = & \by{4 \pi \alpha^2}{s}[e_q^2 + 2e_q v_e v_q \, \rho_1(s)
+(v_e^2 + a_e^2)v_q^2 \, \rho_2(s)]~, \\ 
c_q^A & = & \by{4 \pi \alpha^2}{s}(v_e^2+a_e^2)a_q^2 \rho_2(s)~, \\ \nonumber
\rho_1(s) & = & \by{1}{4 \sin^2 \theta_{\rm w} \cos^2 \theta_{\rm w}}
\by{s(m_Z^2-s)}{(m_Z^2-s)^2+m_Z^2 \Gamma_Z^2}~, \\ \nonumber 
\rho_2(s) & = & \left (\by{1}{4 \sin^2 \theta_{\rm w} \cos^2 \theta_{\rm
w}}\right )^2 \by{s^2}{(m_Z^2-s)^2+m_Z^2 \Gamma_Z^2}~. \nonumber
\end{eqnarray}
In Eq. \ref{eq:totcross}, $T_3$ is the third component of weak isospin,
$\theta_{\rm w}$ is the Weinberg angle, $m_Z, \Gamma_Z$ the mass and
width of the $Z$-boson, and a sum over quarks as well as anti-quarks is
implied. Here $x$ is the energy fraction, $x = E_{\rm hadron}/E_{\rm
beam} = 2 E_h/\sqrt{s}$. Since we neglect the hadron masses, this is
the same as the momentum variable, $x_p = P_{\rm hadron}/P_{\rm beam}
= 2 P_h/\sqrt{s}$; $ x^2 = x_p^2 + 4m_h^2/s$, for a hadron of mass $m_h$
where $Q = \sqrt{s}$ is the energy scale of the interaction. The
fragmentation function, $D_q^h(x,Q^2)$, is the probability at a scale
$Q$ for a quark $q$ to hadronise to a hadron $h$ carrying a fraction $x$
of the energy of the fragmenting quark.

We re-express the cross section in terms of the non-singlet and
singlet fragmentation function combinations, as
\begin{eqnarray}
\by{1}{\sigma^{\rm tot}} \by{\d \sigma^h}{\d x} & = & \by{a_0
D_0^h(x,Q^2) +a_3 D_3(x,Q^2) + a_8 D_8(x,Q^2)}{\sum_q {c_q}} ~,
\label{eq:sns}
\end{eqnarray}
where $D_0$, $D_3$ and $D_8$ refer to the singlet, $D_0 = (D_u + D_d +
D_s)$, and the two non-singlet combinations, $(D_u - D_d)$ and $(D_u
+ D_d - 2 D_s)$, respectively, with $a_0 = (c_u+c_d+c_s)/3$; $a_3 =
(c_u-c_d)/2$ and $a_8 = (c_u+c_d-2c_s)/6$. Again, a sum over both quark
and anti-quark flavours is implied. Note that the $D_0$ term dominates
at the $Z$-pole, when $a_0 \gg a_3, a_8$.

\subsection{Quark versus gluon jet fragmentation}

The OPAL Collaboration \cite{OPAL} also has data separately on charged
hadron production from quark and gluon jets. For the case of charged
hadrons in quark jets, the normalised rate is given by,
\begin{equation}
\left. \by{1}{N} \by{\d N^{\rm ch}}{\d x} \right\vert_q = \by{1}{2}
\sum_h \by{1}{\sigma^{\rm tot}} \by{\d\sigma^h}{\d x}~,
\label{eq:qjet}
\end{equation}
where $1/\sigma^{\rm tot}\, \d\sigma^h/\d x$ is given in
Eq.~\ref{eq:sns} and the sum over charged hadrons $h = \pi^+$, $\pi^-$,
$K^+$, $K^-$, $K_s$ is dominated by the pionic contribution. The factor of
1/2 occurs because the rate per quark jet is measured and there are 2
jets per process. 

The more interesting data is that of charged hadrons from gluon jets.
This arises in 3-jet production via $e^+ e^- \to q \overline{q} g$ where
the gluon jet is isolated by tagging on the (heavy) quark and
anti-quark. The normalised rate here is directly proportional to the
gluon fragmentation function, since the overall kinematical as well as
charged factors cancel:
\begin{equation}
\left. \by{1}{N} \by{\d N^{\rm ch}}{\d x} \right\vert_g = 
\sum_h D_g^h~.
\label{eq:gjet}
\end{equation}
Hence the 3-jet data gives information on the gluon fragmentation
function. 

\subsection{The $p\,p$ process}

In addition to the unknown fragmentation functions, hadro-production in
$p\,p$ processes requires information on the initial state parton density
distributions. The underlying processes here are all possible $q\,q$,
$q\,g$ and $g\,g$ $(2\to 2)$ interactions where one of the final state
partons fragments into the meson of interest.

The invariant inclusive cross section for the reaction $p+p \rightarrow
h+X$ for producing a hadron $h$ at large $p_T$ in the center of mass of
the initial state protons (neglecting the transverse momentum) is given
by~\cite{Reyafrag},
\begin{equation}
E_h\frac{\d^3\sigma}{\d p^3_h} = \frac{1}{\pi} \sum \int_{x_a^{\rm min}}^1
           \d x_a \int_{x_b^{\rm min}}^1 \d x_b\,P_a^A(x_a,Q^2)\,P_b^B(x_b,Q^2)
	   \,\frac{\d\sigma^{ab \rightarrow cd}}{z_h \d \hat{t}} \,
	   D_c^h(z_h,Q^2)~,
\label{eq:pp}
\end{equation}
where the sum over $(a,b,c,d)$ runs over both quarks and gluons. Here
$x_a$ and $x_b$ are the usual Bjorken-$x$ corresponding to the parent
proton momenta $p_A$ and $p_B$: $x_a=p_a/p_A$, $x_b=p_b/p_B$. Hence
$P_{a/A}(x_a, Q^2)$ are the usual parton density distributions; for
example, $P_{u/p}(x_a, Q^2) \equiv u(x_a, Q^2)$, etc.

The fragmentation functions depend on the variables, $z = z_h = p_h/p_c$
and $Q^2 = p_T^2$ while the partonic sub-process variables are as
usual defined in terms of the $s$, $t$ and $u$ hadronic variables as
$\hat{s}=x_a x_b s$, $\hat{t}=x_a t / z$ and $\hat{u}=x_b u / z$. The
limits of integration are~\cite{Reyafrag}
\begin{equation}
\label{eq:xmin}
x_a^{min} = \frac{x_1}{1-x_a}~;~~
x_b^{min} = \frac{x_a x_2}{x_a-x_1}~, \nonumber
\end{equation}
with $x_1=-u/s, x_2=-t/s$.

For numerical comparison with the data, we re\"express the
cross-section in terms of the physical observables which are the
transverse momentum $p_T = p_h \sin\theta$ and the rapidity
$y=(1/2)\ln [(E_h + p_h \cos\theta)/(E_h - p_h \cos\theta)]$, as
\begin{equation}
E_h\frac{\d^3\sigma}{\d p^3_h} \equiv
\frac{1}{2p_T}\frac{\d^3\sigma}{\d p_T \d y \d\phi}~,
\label{eq:pTy}
\end{equation}
where $\theta$ is the
scattering angle of the hadron $h$ in the $p\,p$ center of mass frame
and $E_h$ and $p_h$ are its energy and 3-momentum. 


The various sub-process cross-sections \cite{CM} are listed in
Table~\ref{tab:pphat}. Note that the $q\,q$, $q\,g$ and $g\,g$ processes
all contribute at the same order in $\alpha_s$. Hence the quark and
gluon fragmentation functions contribute at the same order, unlike in
the $e^+\,e^-$ case.

\begin{table}[htb]
\centering
\begin{tabular}{|c|c|} \hline 
Subprocess & $\frac{d\sigma^{ab \rightarrow hd}}{d \hat{t}}\frac{1}{z_H}$\\ [2ex] \hline
$q_i q_j \rightarrow q_i q_j,$ & $2F(\chi) + 0$ \\
$q_i \bar{q_j} \rightarrow q_i \bar{q_j}$ & \\[2ex] \hline
$q_i q_i \rightarrow q_i q_i$ & $2F(\chi) - \frac{2}{N}\{ \chi+2+\frac{1}{\chi} \}$\\[2ex] \hline
$q_i \bar{q_i} \rightarrow q_i \bar{q_i}$ & $2F(\chi) + \{\frac{2}{N}(\chi-1+\frac{1}{\chi})+\frac{2(1+\chi^2)}{(1+\chi)^2} \}$ \\
$q_i \bar{q_i} \rightarrow q_j \bar{q_j}$ & $0 + \{\frac{2(1+\chi^2)}{(1+\chi)^2} \}$ \\
$q_i \bar{q_i} \rightarrow gg$ & $0 + \frac{N_g}{N}\{\chi + \frac{1}{\chi} - \frac{h(1+\chi^2)}{(1+\chi)^2} \}$\\[2ex] \hline
$q g \rightarrow q g,$ & $h 2 F(\chi) + \{ \chi +3 + \frac{1}{\chi} \}$\\
$\bar{q} g \rightarrow \bar{q} g $ & \\[2ex] \hline
$ g g \rightarrow g g $ & $ h^2 2 F(\chi) + h^2 \{4 - \frac {2\chi}{(1+\chi)^2} \} $\\
$ g g \rightarrow q \bar{q} $ & $0 + \frac{N}{N_g}\{2(\chi + \frac{1}{\chi}) - \frac{ 2 h(1+\chi^2)}{(1+\chi)^2} \}$\\[2ex] \hline
\end{tabular}
\caption{Subprocess cross-sections in $p\,p$ scattering \cite{CM} in units of
$a \pi \alpha_s^2/{\hat{s} }^2$; $a=4C_F^2/N_g, h=C_A/C_F$,
with $C_F = 4/3$, $C_A = 3$, $N=3$, and $N_g = 8$.}
\label{tab:pphat}
\end{table}

We now present details of our model for quark fragmentation functions.

\section{The Model} 

This model was developed in Ref.~\cite{IMR} to study $\pi$ and $K$
fragmentation in $e^+\,e^-$ collisions. Consider inclusive octet
hadro-production under the assumption of exact SU(3) (flavour) symmetry:
\begin{equation}
\label{eq:qhx}
q_i \rightarrow h^j_i +X_j~,
\end{equation}
where $q_i = u, d, s$ for $i = 1, 2, 3$. That is, we have $3 \to 8
+ X$, with $X$ being a triplet,
antisixplet or fifteenplet. This holds for both octet meson and baryon
production.  The fragmentation of a light quark ($u$, $d$, $s$) into
any member of the octet is completely described by three independent
fragmentation functions, $\alpha(x,Q^2)$, $\beta(x,Q^2)$, $\gamma(x,Q^2)$,
corresponding to $X=3,\bar{6},15$ respectively.

The corresponding fragmentation functions for the antiquarks are
$\bar{\alpha}$, $\bar{\beta}$ and $\bar{\gamma}$ respectively. As
the pseudo-scalar meson-octet contains both the mesons and their
antiparticles, $D^h_q(x, Q^2)=D^{\bar{h}}_{\bar{q}}(x, Q^2)$ so that
there are only three independent quark fragmentation functions for the
entire meson octet, which we choose to be $\alpha$, $\beta$ and $\gamma$.

These simplify further on applying equality of unfavoured fragmentation
so that $D_u^{\pi^-}=D_d^{\pi^+}=D_s^{\pi^+}=D_s^{\pi^-}$, etc.,
which reduces the number of independent functions to a valence
fragmentation function, $V(x, Q^2)$ and a sea fragmentation function,
$S(x, Q^2)$, where
\begin{eqnarray}
\label{eq:ff}
V & = & \alpha+\beta-\frac{5}{4}\gamma=\alpha-\frac{3}{4}\gamma~; \\
S & = & 4\beta = 2\gamma~.
\end{eqnarray}
The detailed expressions for $D_q^h$ in terms of these functions are given
in Ref.~\cite{IMR} and are reproduced in Table~\ref{tab:frag}
in terms of $V$ and $\gamma$.

\begin{table}[htb]
\centering
\begin{tabular}{|ccl|ccl|} \hline
fragmenting & \multicolumn{2}{c|}{${}_{\displaystyle p/K^+}$} & fragmenting &
\multicolumn{2}{c|}{${}_{\displaystyle n/K^0}$} \\
quark & & & quark & & \\  \hline
$u$ & : &  ${V + 2\gamma}$ & 
$u$ & : &  $2{\gamma}$ \\
$d$ & : &  $2{\gamma}$ & 
$d$ & : &  ${V + 2\gamma}$ \\
$s$ & : &  $2 {\gamma}$ & 
$s$ & : &  $2 {\gamma}$ \\ \hline
fragmenting & \multicolumn{2}{c|}{${}_{\displaystyle \Lambda^0/\eta}$} & fragmenting &
\multicolumn{2}{c|}{${}_{\displaystyle \Sigma^0/\pi^0}$} \\
quark & & & quark & & \\  \hline
$u$ & : &  
$\frac{1}{6}{V}+{2\gamma}$ &
$u$ & : &  
$\frac{1}{2}{V} + 2{\gamma}$ \\
$d$ & : &  
$\frac{1}{6}{V}+2{\gamma}$ &
$d$ & : &  
$\frac{1}{2}{V}+2{\gamma}$ \\
$s$ & : &  
$\frac{4}{6}{V}+2{\gamma}$ &
$s$ & : &  $2\gamma$ \\ \hline
fragmenting & \multicolumn{2}{c|}{${{}_{\displaystyle \Sigma^+/\pi^+}}$} &
fragmenting &  \multicolumn{2}{c|}{${{}_{\displaystyle \Sigma^-/\pi^-}}$} \\
quark & & & quark & & \\  \hline
$u$ &  : & ${V}+{2\gamma}$ & 
$u$ &  : & $2 {\gamma}$ \\ 
$d$ &  : & $2 {\gamma}$ & 
$d$ &  : & ${V}+2{\gamma}$ \\ 
$s$ &  : & $2{\beta}+{\gamma}$ & 
$s$ &  : & $2{\gamma}$ \\ \hline
fragmenting & \multicolumn{2}{c|}{${{}_{\displaystyle \Xi^0/\overline{K^0}}}$} &
fragmenting & \multicolumn{2}{c|}{${{}_{\displaystyle \Xi^-/K^-}}$} \\
quark & & & quark & & \\  \hline
$u$ & : & $2{\gamma}$ & 
$u$ & : & $2 {\gamma}$ \\ 
$d$ & : & $2 {\gamma}$ & 
$d$ & : & $2{\gamma}$ \\ 
$s$ & : & ${V}+2{\gamma}$ & 
$s$ & : & ${V}+2{\gamma}$ \\
\hline
\end{tabular}
\caption{Quark fragmentation functions into members of the
baryon and meson octet in terms of the SU(3) functions $V(x)$
and $\gamma(x)$ in the exact SU(3) symmetric case. }
\label{tab:frag}
\end{table}

\subsection{Flavour breaking effects}

Flavour SU(3) is badly broken due to the relatively heavier $s$-quark
compared to the $u$- and $d$-quarks. This is implemented in the model
through an $x$-independent suppression factor $\lambda$ whenever a
strange quark is needed in order to fragment into that meson. For example,
$D_u^{K^+}$, $D_d^{K^+}$ and $D_s^{K^+}$ are suppressed by $\lambda$
compared to the SU(3) symmetric expressions given in Table~\ref{tab:frag}
since we require $q \to q\overline{s}$ in order to fragment
into $K^+$ with a valence $\overline{s}$. While $D_{\overline{u}}^{K^+}$ and
$D_{\overline{d}}^{K^+}$ are similarly suppressed, only the sea part of
$D_{\overline{s}}^{K^+}$ is suppressed, so that the fragmentation of
$\overline{s}$ into $K^+$ is given by (see Table~\ref{tab:frag}),
\begin{equation}
D_{\overline{s}}^{K^+} = 2V + 2 \lambda \gamma~.
\end{equation}
The valence component is not suppressed since the heavier strange quark
is already available and only non-strange quarks are needed to produce
$K^+$. The sea (anti)-quark fragmentation function is suppressed since,
by definition, the fragmenting (anti)-quark is in the sea of the
produced hadron and a strange (anti)-quark is still needed to make up
the valence quantum numbers. Similar arguments can be applied to the
fragmentation functions of $K^-$, $K^0$ and $\overline{K^0}$.

In short, the sea contributions to the $K$ meson fragmentation functions
remain SU(3) symmetric and are uniformly suppressed by $\lambda$ compared
to the corresponding unbroken SU(3) (or $\pi$) fragmentation functions.
Hence, all the quark fragmentation functions, and hence
the combinations $D_0$, $D_3$ and $D_8$ are expressed in terms of the
octet functions, $V$ and $\gamma$ and the constant suppression factor
$\lambda$, which we expect to be $\lambda \sim m_\pi^2/m_K^2$.

Until now nothing new has been introduced. We now extend this model to
include the singlet mesons and SU(3) singlet--octet mixing, in order to
determine the $\eta$ and $\eta'$ fragmentation functions.

\subsection{Extension for $\eta,\eta^{\prime}$ mesons }

The physical $\eta$ and $\eta'$ states are orthogonal admixtures of the
pure SU(3) octet and singlet states, $\eta_8$ and $\eta_1$, with mixing
angle $\theta_P$:
\begin{eqnarray}
\eta&=& \eta_8 \cos\theta_P - \eta_1 \sin\theta_P~;\nonumber\\
\eta^{\prime}&=& \eta_8 \sin\theta_P + \eta_1 \cos\theta_P~,\nonumber
\label{etaetap}
\end{eqnarray}
where $\vert \eta_8\rangle = (u\overline{u} + d\overline{d}-2
s\overline{s})/\sqrt{6}$ and
$\vert \eta_1\rangle = (u\overline{u} + d\overline{d} +
s\overline{s})/\sqrt{3}$.
The quadratic and linear mass formulas predict $\theta_P$ to be $-11.5^\circ$
and $-24.6^\circ$ respectively; present experimental limits allow
the range $-24^\circ \le \theta_P \le -10^\circ$ \cite{pdg}. Note that
$\theta_P$ is still small; $\cos\theta_P \ge 0.9$, and so $\eta$ ($\eta'$) is
dominated by its $\eta_8$ ($\eta_1$) component.

The purely octet contributions are given in Table~\ref{tab:frag}. Note
that the singlet combination of the $\eta_8$ fragmentation function
is given by $D_0^8 = 2V + 12 \gamma$ just as for the $\pi$ and $K$
mesons. The non-singlet fragmentation functions are $D_3^8 = 0$ and
$D_8^8 = -2V$.

\subsection{SU(3) singlet fragmentation functions}

In order to complete the description for the physical $\eta$ meson,
we now need to address the SU(3) singlet contributions. There is only
one singlet fragmentation function that we denote as $\delta$, since
the only possibility is $3 \to 1 + X$:
\begin{equation}
\label{eq:qhx1}
q_i \rightarrow h +X_i~,
\end{equation}
where $q_i = u, d, s$ for $i = 1, 2, 3$, with $X$ being a triplet alone.
Instead of including an additional unknown fragmentation function, we
follow a slightly different route. Since $\delta$ describes a pure SU(3)
singlet fragmentation function, it should be proportional to the
singlet fragmentation function $D_0^8$. We set
\begin{equation}
\delta \equiv D_0^1 = f_d (D_0^8) = f_d (2V + 12\gamma)~,
\end{equation}
where we choose $f_d$ to be an $x$-independent (unknown) constant.
For the same reason, we set the non-singlet functions $D_3^1 = D_8^1 =
0$. These conditions completely determine the individual $\eta_1$ quark
fragmentation functions in terms of $f_d$ as given in
Table~\ref{tab:singlet}.

\begin{table}[htb]
\centering
\begin{tabular}{|ccl|} \hline
fragmenting & \multicolumn{2}{c|}{${}_{\displaystyle \eta_1}$} \\
quark & & \\  \hline
$u$ & : &  $f_d (\frac{1}{3} {V} + 2{\gamma}$) \\
$d$ & : &  $f_d (\frac{1}{3} {V} + 2{\gamma}$) \\
$s$ & : &  $f_d (\frac{1}{3} {V} + 2{\gamma}$) \\ \hline
\end{tabular}
\caption{The unbroken SU(3) singlet $\eta_1$ fragmentation functions;
$f_d$ is a free parameter.}
\label{tab:singlet}
\end{table}

\subsection{Mass suppression effects in the $\eta_8$ fragmentation functions}

\subsubsection{Valence sector}

We now consider SU(3) breaking effects and proceed along the lines
discussed for suppression in $K$ mesons. Since $\eta_8$ is a bound state
of $q_i\overline{q}_i$ quarks of the same flavour, only the strange
quark valence fragmentation functions, i.e., the valence components
of $D_s^8$ and $D_{\overline{s}}^8$ are suppressed; this
suppression is again by the same factor $\lambda$ as in the kaon case.

\subsubsection{Sea sector}

There is an additional complication for the sea quark fragmentation
relative to the kaon sector: a sea quark has to pick up a
quark--anti-quark pair of the {\em same} flavour so that $q_i
\overline{q}_i \to \eta_8$, but all possible flavours, $i = u, d, s$,
are possible because of the flavour content of the $\eta_8$ meson. If a
strange quark--anti-quark pair is produced during fragmentation, the
contribution is suppressed by $\lambda$, exactly as in the case of kaons.

A priori, it appears that there should be no suppression when
$u\overline{u}$ or $d\overline{d}$ pairs fragment into $\eta_8$. However,
a non-strange $u \overline{u}$ or $d \overline{d}$ pair will prefer to
fragment to $\pi$ rather than $\eta$ because of the relative masses
involved. Hence, non-strange sea fragmentation to $\eta_8$ is also
suppressed; since the strangeness suppression factor is $\lambda =
m_\pi^2/m_K^2$, we take the non-strange suppression factor to be
$\lambda_8 = m_\pi^2/m_\eta^2$. Note that we have used the relevant mass
as that of the physical mass eigenstate, $m_\eta$, since the $\eta_8$
contribution dominates the $\eta$ state. Since sea quark contributions to
the $\eta_8$ fragmentation can arise from production of quark--anti-quark
pairs of any flavour, hence they are uniformly suppressed by a factor
$f_8 = (2\lambda_8 + \lambda)/3$.

\subsection{Mass suppression factors in $\eta_1$ fragmentation functions}

The logic is exactly the same as for the $\eta_8$ case. The valence strange
sector is suppressed by $\lambda$ as before while the sea sector is
uniformly suppressed by a factor $f_1 = (2\lambda_1 + \lambda)/3$ where
$\lambda_1 = m_\pi^2/m_{\eta'}^2$ is the non-strange suppression factor
for $\eta_1$, assuming that $\eta_1$ dominates the $\eta'$ state.

\subsection{The $\eta$, $\eta'$ fragmentation functions} Hence, the model
has been extended to include a realistic prediction of the $\eta_8$ and
$\eta_1$ fragmentation functions in terms of the two octet functions $V$
and $\gamma$ and the $x$-independent suppression factor $\lambda$ (all
known from fits to $\pi$ and $K$ data), along with the $x$-independent
suppression factors $\lambda_8$ and $\lambda_1$ apart from an unknown
constant $f_d$.

In summary, we have, for the octet meson:
\begin{eqnarray}
\label{eq:frageta8}
D_u^8 = D_d^8 &=& \frac{1}{6}V + 2f_8 \gamma~,\nonumber\\
D_s^8 &=& \frac{2}{3}V\lambda + 2f_8\gamma~,\nonumber
\end{eqnarray}
and for the singlet meson:
\begin{eqnarray}
\label{eq:ufrageta1}
D_u^1 = D_d^1 &=& f_d\left(\frac{1}{6}V + 2f_1 \gamma \right)~,\nonumber\\
\label{eq:sfrageta1}
D_s^1 &=& f_d\left(\frac{2}{3}V\lambda + 2f_1 \gamma\right)~.\nonumber
\end{eqnarray}

The fragmentation functions for the physical states can then be
expressed in terms of the fragmentation functions for the $\eta_8$ and
$\eta_1$ parts, along with the mixing angle, $\theta_P$. We have
\begin{eqnarray}
D_i^\eta & = & (c_i^\eta)^2 \left({\cos^2\theta_P}\by{D_i^8}{(c_i^8)^2} + 
{\sin^2\theta_P}\by{D_i^1}{(c_i^1)^2} \right)~, \\
D_i^{\eta'} & = & (c_i^{\eta'})^2 \left({\sin^2\theta_P}\by{D_i^8}{(c_i^8)^2} + 
{\cos^2\theta_P}\by{D_i^1}{(c_i^1)^2} \right)~,
\end{eqnarray}
where $i = u, d, s$ and the coefficients are $c_u^8=c_d^8=1$, $c_s^8 =
-2$, $c_u^1=c_d^1=c_s^1=\sqrt{2}$, $c_u^\eta =c_d^\eta=
(\cos\theta_P-\sqrt{2}\sin\theta_P)$, and $c_s^\eta = 
(-2\cos\theta_P-\sqrt{2}\sin\theta_P)$. The coefficients for $\eta'$ are
obtained from those of $\eta$ by the substitution: $(\cos\theta_P \to
\sin\theta_P; \sin\theta_P \to -\cos\theta_P)$.

\section{Comparison with the data}

We now fit the unknown model parameters by a comparison with data. Note
that the octet fragmentation functions $V(x, Q^2)$ and $\gamma(x, Q^2)$,
the suppression factor, $\lambda$, and the gluon fragmentation function
$D_g(x, Q^2)$, have already been fitted to the $\pi$ and $K$ data in
Ref.~\cite{IMR}. We consider them here again for reasons explained
below.

\subsection{$e^+e^- \to$ 2-jets}

The $V$ and $\gamma$ were fitted to parametrised forms as a function of $x$:
\begin{equation}
F_i(x) = a_i(1-x)^{b_i}(x^{c_i})(1+d_ix+e_ix^2)~,
\end{equation}
at a starting scale of $Q_0^2 = 2$ GeV${}^2$ and evolved, along with
the gluon fragmentation function, $D_g(x, Q^2)$, to leading order (LO)
in perturbative QCD. Best fits to $V(x, Q_0^2)$, $S(x, Q_0^2)$, $D_g(x,
Q_0^2)$ and $\lambda$ were obtained from fits to electro-production data
at the $Z$-pole $Q^2 = 91.2^2$ GeV$^2$ for $\pi^{\pm}$ and $K^{\pm}$
\cite{Lafferty}.

The best-fit values for the parameters $a, b, c, d, e$ for different
input fragmentation functions are given in Table~\ref{tab:inputll}. The
best fit value for $\lambda$ is 
\begin{equation}
\lambda = 0.08~,
\end{equation}
which is consistent with the quark model expectation $\lambda =
m_\pi^2/m_K^2$.

\begin{figure}[htp]
\centering
\includegraphics[trim=0 40 0 260, width=\textwidth,clip]{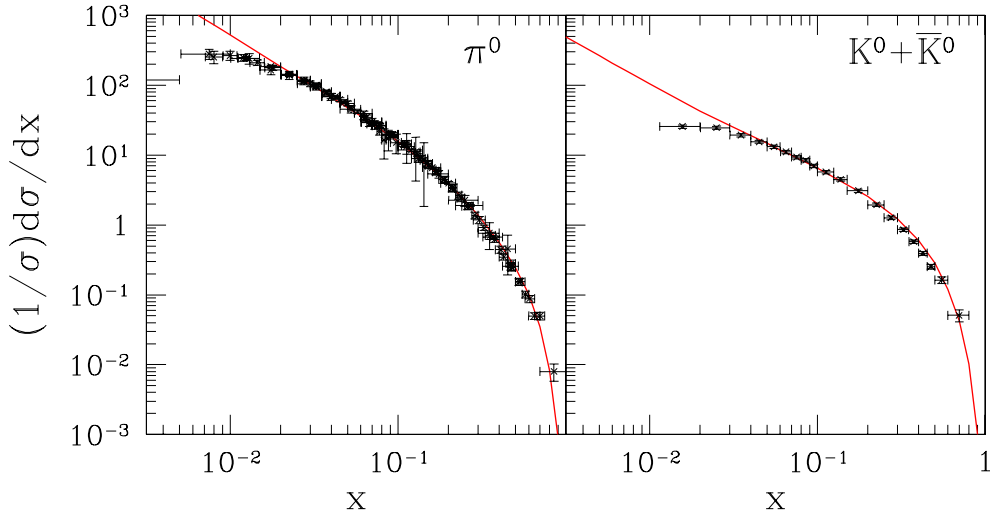}
\caption{The model fits to the $Z$-pole data for $\pi^0$ \cite{LEPpi} and
$K^0+\overline{K}^0$ \cite{LEPK} mesons; the functions at the starting scale
$Q_0^2 = 2$ GeV$^2$ are given in Table~\ref{tab:inputll}.}
\label{fig:piK}
\end{figure}

The resulting fits to the $\pi$ and $K$ production rates at the $Z$-pole
from LEP data \cite{LEPpi,LEPK} are shown in Fig.~\ref{fig:piK}.
Note that the gluon is not well-determined
from this data, unlike the quark fragmentation functions. This is
because the cross-sections are functions only of the quark fragmentation
functions at LO and the gluon contribution only enters through the
evolution equations. The electro-production rates are insensitive to the
gluon even at next-to-leading order (NLO) since the gluon contribution
is down by a factor of $\alpha_s$ compared to the quark contributions.
Three-jet processes are sensitive to the gluon contribution; we return
to this point later.

Note that the sea quark and gluon distributions are different from
those obtained in Ref.~\cite{IMR}. Since gluons and
sea quarks mix during singlet evolution, a small non-zero gluon was
included earlier while evolving from the starting scale $Q_0^2 = 2$
GeV$^2$ to the observed scale $Q^2$. Gluon fragmentation function data
is now available \cite{OPAL} from isolating charged hadrons produced from
gluon jet fragmentation in $e^+\,e^- \to q \overline{q} g$ processes. It
turns out that the gluon fragmentation used in Ref.~\cite{IMR} is too
small and therefore incompatible with the 3-jet data. Our new updated
fits here include a gluon that is consistent with such data (as discussed
below); this necessitates a revision of the sea quark {\em starting}
fragmentation functions (they have been made smaller) so that the sea
contribution evolved to the $Z$-pole (with the contribution from the
new enhanced gluon) still matches the earlier result (and hence the data).

The valence function $V(x, Q_0^2)$ remains the same as before.

It is observed that the predictions exceed the data systematically at
small-$x$. This is because of the pole in the $P_{gq}$ and $P_{gg}$
splitting functions as $x\to 0$ which cause the DGLAP evolution to drive
the small-$x$ sea and gluon fragmentation functions to large valueswhich
cause the DGLAP evolution to drive the small-$x$ sea and gluon
fragmentation functions to large values. While the starting values of
the fragmentation functions are integrable, this evolution makes them
non-integrable due to this increase at small-$x$. However, the
fragmentation functions should remain integrable (as their first moments
are related to the hadron multiplicities). An approach such as the
modified leading log approximation (MLLA) \cite{Dok} cures the small-$x$
divergence and gives a good fit to the data at small-$x$ as has been
discussed in Ref.~\cite{IMR}; here we simply ignore the small-$x$
region and concentrate on the fits for $x \gtrsim 0.02$ for all mesons
under consideration.

\begin{table}[htb]
\centering
\begin{tabular}{|c|c|c|c|} \hline
      & {$2V$} & {$2\gamma$} & {$2g$} \\ \hline
$a$ & 2.33 & 3.5 & 7.25 \\ \hline
$b$ & 2.15 & 12.76 & 4.4 \\ \hline
$c$ & -0.64 &-0.75 & $-0.5$ \\ \hline
$d$ & 5.35 & 3.87 & 2 \\ \hline
$e$ & -5.12 & 61.59 & 0 \\ \hline
\end{tabular}
\caption{Input values at $Q^2 = 2$ GeV${}^2$ for the valence, sea
and gluon fragmentation functions for the pseudo-scalar meson octet. The
factor of 2 is due to the convention used in the earlier work \cite{IMR}
when the model was fitted to the {\em sum} of the charged hadrons, for
example, $\pi^++\pi^-$, $K^++K^-$, etc.}
\label{tab:inputll}
\end{table}

\subsubsection{Fits to $\eta$, $\eta'$ data on the $Z$ pole}

\begin{figure}[thp]
\centering
\includegraphics[trim=0 40 0 260,width=\textwidth,clip]{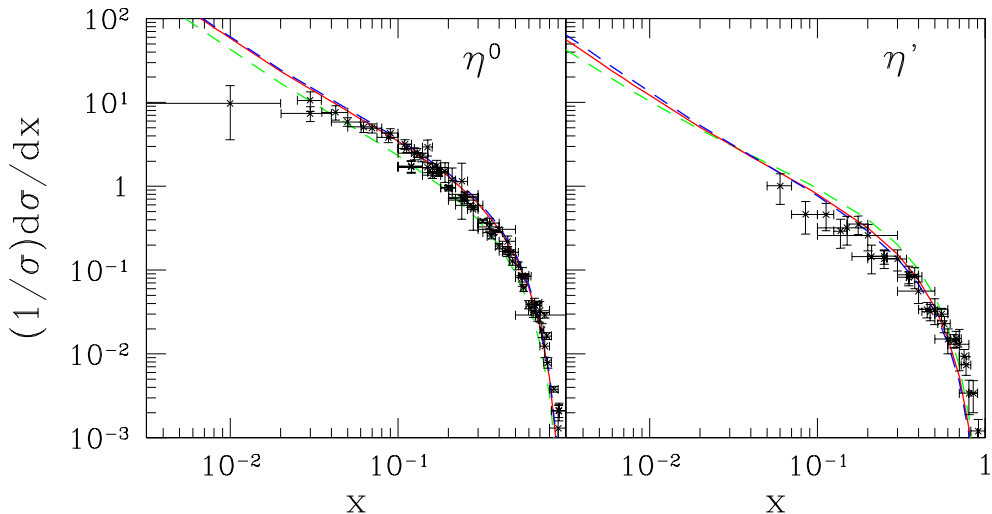}
\caption{The figure shows the model fits at $\sqrt{s} = 91.2$ GeV to
the LEP data on $\eta$ \cite{LEPeta} and $\eta^{\prime}$ \cite{LEPetap}
mesons as a function of $x$ with both $x$ and $y$ error-bars. The green,
red and blue lines represent $\theta_P=0^\circ,-15^\circ$ and $-24^\circ$
respectively for $f_d = 0.3$.}
\label{fig:etaetap}
\end{figure}

With $\lambda_8$ and $\lambda_1$ fixed to the theoretically expected
values, $f_d$ is the only unknown parameter in the rates for both $\eta$
and $\eta'$ production. Since the nonet mixing angle $\theta_P$ is
small, $f_d$ is mostly constrained by the large-$x$ data. Fits to the
$\eta$, $\eta'$ production rates in $e^+$-$e^-$ annihilation on the
$Z$-pole at LEP \cite{LEPeta,LEPetap} yield a best fit value to the free parameter
of $f_d=0.3$. However, the electro-production data are insensitive to
the nonet mixing angle $\theta_P$, as can be seen from the fits to the
data in Fig.~\ref{fig:etaetap} where the fits for $\theta_P = 0^\circ,
-15^\circ$ and $-24^\circ$ are shown. The allowed range of $\theta_P$
marginally alters the fits at small-$x$ values, while being completely
insensitive in the large-$x$ region.

\subsection{$p\,p \to (\pi, \eta) X$}

The PHENIX experiment at RHIC \cite{phenixpi, phenixeta} has measured
the inclusive transverse momentum spectra of $\pi$ and $\eta$ mesons in
the range $p_T \approx 1.5$--14 GeV$/c$ at mid-rapidity $(|\eta| < 0.35)$
in $p\,p$, $d\,Au$ and $Au\,Au$ collisions at $\sqrt{s_{NN}}=200$
GeV. They have also measured the $\eta/\pi^0$ production ratio,
$R_{\eta/\pi^0}$. This ratio is interesting because of its potential as
a good signal for quark-gluon-plasma.

The $\eta/\pi^0$ ratio in $p\,p$ collisions is the base-line with which
comparisons in $d\,A$ or $A\,A$ processes are made. It is therefore
important to understand and correctly determine this ratio in $p\,p$
collisions for a wide range of $Q^2$ ($\sim p_T^2$). As can be seen from
Eq.~\ref{eq:pp}, this depends on the initial parton density
distributions and the final state fragmentation functions. Furthermore,
the gluon and quark fragmentation functions (and density distribution
functions) occur at the same order in $\alpha_s$. The study of such a
process, therefore, may help determine the gluon fragmentation functions
with much greater accuracy than currently known. With these two
considerations in view, the model was compared to the $\eta$ and
$\pi$ production rates in $p\,p$ collisions.

\subsubsection{Analytical approximation}

Before comparing the model to the data, it is observed that there is
considerable simplification of the expressions when the $p\,p$
sub-process cross-sections are re\"expressed as a constant times a
kinematic factor, $F(\chi)$, with small correction terms \cite{CM}, as
can be seen from Table~\ref{tab:pphat}, where
\begin{equation}
\label{eq:fchi}
F(\chi) = \chi^2 + \chi + 1 + \frac{1}{\chi} + \frac{1}{\chi^2},
\end{equation}
where $\chi = \hat{u}/\hat{t}$. When these correction terms are dropped,
the surviving terms are such that the fragmenting parton $c$ has the
same flavour as the initial parton $a$. Then, apart from an overall
factor,
\begin{equation}
f = \by{C_F^2}{N_g} \by{4\pi\alpha_s^2}{\hat{s}^2}~,
\end{equation}
the integrand of Eq.~\ref{eq:pp} for the hadro-production rate can be
factorised as the product of two terms:
\begin{equation}
\sum_{i,j} \left(q_i D_i^h+ \overline{q}_i \overline{D}_i^h +
     h \, g D_g^h \right) \times
\left(q_j + \overline{q}_j + h \, g\right)~.
\end{equation}
The first term depends on $(x_a, z)$ and contains the dependence on
the unknown fragmentation functions while the second term depends on
$x_b$ alone. In this factorised form it is clear that there is a
substantial dependence on the gluon fragmentation function only when the
corresponding gluon density distribution function is large as well; this
occurs, for the central rapidity region measured by PHENIX at RHIC, for
low $p_T$, $p_T \lesssim 5$ GeV. Hence, the small-$p_T$ data
(with $Q^2 \approx p_T^2 \ge 2$ GeV$^2$) should be sensitive to the gluon
fragmentation function. Since our input fragmentation functions are at a
scale $Q_0^2 = 2$ GeV$^2$, there may be significant threshold effects at
this energy; we choose $Q^2 = (p_T^2 + m_h^2)$ with $Q^2 > Q_0^2$.

\subsection{Fits to the $\pi$ production rate}

\begin{figure}[htp]
\begin{center}
\includegraphics[trim = 0 25 260 280,width=0.65\textwidth,clip]{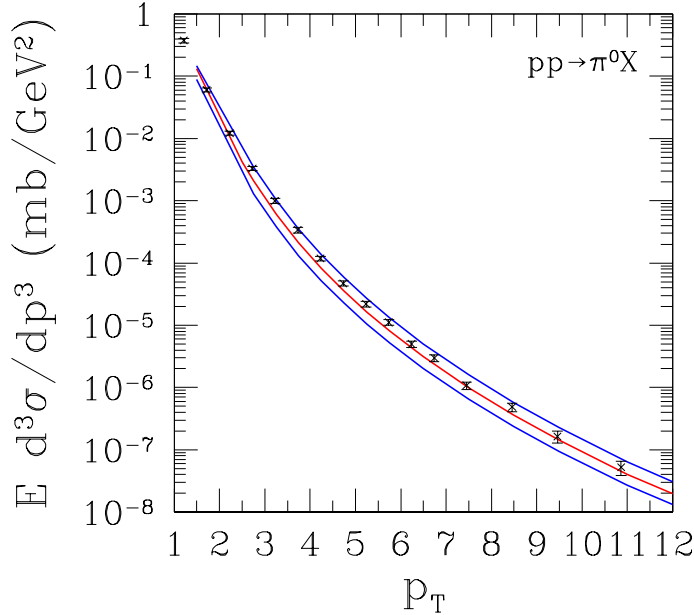}
\end{center}
\caption{The $\pi$ production rate in $p\,p$ collisions at $\sqrt{s} =
200$ GeV in comparison with the RHIC/PHENIX data \cite{phenixpi}. The
central curve corresponds to $Q^2 = p_T^2+m_h^2$ while the upper
(lower) curve corresponds to a value of $Q^2$ half (twice) this value.}
\label{fig:pi}
\end{figure}

The expressions for the production rate are integrated over the
central rapidity region, $-0.35 \le y \le 0.35$, and over the allowed
ranges of $x_a$ and $x_b$ for a given $p_T$.  The fits to the $\pi$
production data \cite{phenixpi} as a function of $p_T$ are shown in
Fig.~\ref{fig:pi}. Here the exact expressions for the sub-process
cross-sections have been used, not the approximations discussed in the
previous sub-section. The small-$p_T$ data, as expected (where $z_h
\gtrsim 0.02$), are very sensitive to the gluon fragmentation function,
while the larger $p_T$ data ($z_h \gtrsim 0.1$) are sensitive to the quark
fragmentation functions. The effect of scale, $Q^2 = f_S (p_T^2+m_h^2)$,
$f_S=0.5$--2, is also shown in the figure.

\subsection{Fits to the $\eta$ production rate}

It is reasonable to assume that the gluon fragmentation function in
$\eta$ is suppressed relative to that of $\pi$, just as is the case with
sea quarks. However, there is a difference during evolution: the gluon
mixes with the entire singlet combination, $D_0^\eta$, which contains
both valence and sea quark terms. For example, 
\begin{eqnarray} \nonumber
D_0^\pi & = &  2V + 12\gamma~; \\ \nonumber
D_0^K & = &  \by{(1+\lambda)}{2} \, 2V + \lambda \, 12\gamma~; \\
D_0^\eta & = &   x_0 \, 2V + y_0 \, 12\gamma~; \nonumber
\end{eqnarray}
where
\begin{eqnarray} \nonumber
x_0 & = & \by{1}{6} \, \left[2(c_u^\eta)^2 + \lambda (c_s^\eta)^2\right]
\, \left({\cos^2\theta_P} + {\sin^2\theta_P} \right)~, \\
 & \stackrel{\theta_P \hbox{ small}} {\to} & \sim 0.5~; \\ \nonumber
y_0 & = & \by{f_8}{12} \left({\cos^2\theta_P} \, \left[8(c_u^\eta)^2 +
(c_s^\eta)^2 \right] \, + f_d f_1 {\sin^2\theta_P} \,
\left[4(c_u^\eta)^2 + 2(c_s^\eta)^2 \right]\right)~, \\
 & \stackrel{\theta_P \hbox{ small}} {\to} & f_8~. \nonumber
\end{eqnarray}
Hence the valence in $D_0^{K, \eta}$ is suppressed by a factor of 1/2
and the sea is suppressed by $\lambda, f_8$ compared to
$D_0^\pi$. Therefore, we expect the gluon fragmentation into $\eta$ to be
suppressed by $D_g^\eta \sim D_g^K = f_g D_g^\pi$, where $y_0 \le f_g
\le x_0$.

Since $V$ and $\gamma$ dominate in different regions of phase-space (at
different $z_h$ values), it is likely that $f_g$ is an $x$-dependent
suppression factor. This can be studied in 3-jet processes in $e^+\,
e^-$ collisions (see the next sub-section). Since detailed data on this
is still not available, we assume $f_g$ to be an $x$-independent
constant between $y_0$ and $x_0$; certainly $f_g \le 1$.

\begin{figure}[htp]
\begin{center}
\includegraphics[trim = 0 10 0 350, width=\textwidth,clip]{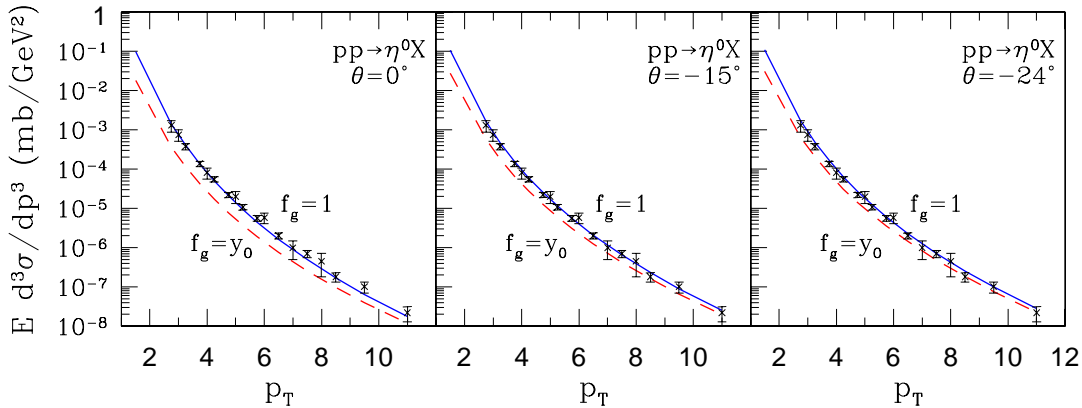}
\end{center}
\caption{The $\eta$ production rate in $p\,p$ collisions at $\sqrt{s} =
200$ GeV as a function of $p_T$ with $Q^2 = p_T^2+m_h^2$
in comparison with the RHIC/PHENIX data \cite{phenixeta}.}
\label{fig:eta}
\end{figure}

The resulting fits to the $\eta$ production rate in $p\,p$ collisions
are shown in Fig.~\ref{fig:eta} for different allowed mixing angles
$\theta_P$. While the data is sensitive to $f_g$, as expected, it does
not appear to be very sensitive to $\theta_P$. Note that the maximum
separation between the curves for a given mixing angle is at
small-$p_T$, where the gluon contributes maximally. At large enough
$p_T$, the ratio should determine $\theta_P$ as the gluon contribution
becomes negligible.

\subsection{Fits to the $\eta/\pi^0$ ratio}

\begin{figure}[thp]
\begin{center}
\includegraphics[trim = 0 10 0 350, width=\textwidth,clip]{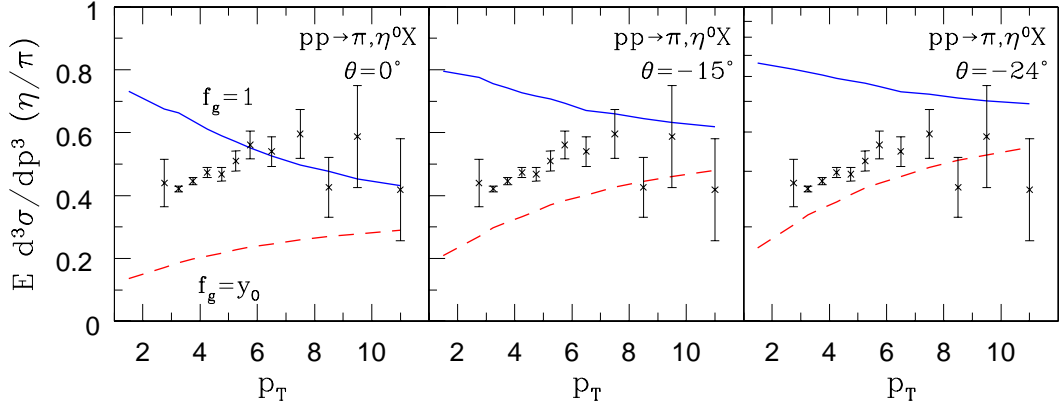}
\end{center}
\caption{The $\eta/\pi$ production ratios in $p\,p$ collisions as a
function of $p_T$ for different nonet mixing angle $\theta_P$ for two
different gluon suppression factors, $f_g = y_0, 1$, in comparison with
the RHIC/PHENIX data at $\sqrt{s} = 200$ GeV \cite{phenixeta}}.
\label{fig:etapi_all}
\end{figure}

The sensitivity to both $f_g$ and $\theta_P$ is amplified in the
$\eta/\pi$ ratio, as can be seen in Fig.~\ref{fig:etapi_all} where the
RHIC/PHENIX data \cite{phenixeta} are also superposed. The region bounded
by the curves for $f_g = y_0, 1$ in the figure encompass the possible
fits to the ratio.

A $\chi^2$ minimisation gives a best fit for $f_g=0.3, \theta_P =
-20.0^\circ$. Since the total errors are large, only the statistical errors
were used in the fit to give $\chi^2 = 9.5$ for 11 d.o.f. A fit with the
total errors reduces this best fit value to $\chi^2 = 4.5$. The saturation
value of the $\eta/\pi$ ratio increases with increasing $\vert
\theta_P\vert$. Constraining the large $p_T$ ratio to be $\sim 0.5$
\cite{phenixeta} as favoured by PYTHIA \cite{PYTHIA} decreases the value
of $\theta_P$ to about $-16^\circ$; however, the errors on the data also
allow this ratio to saturate at values of around 0.6. An improvement in the
data quality will definitely constrain both $\theta_P$ and $f_g$ better.
Currently, a reasonable quality of fit ($\chi^2 \sim$ d.o.f.) is obtained
in a range $f_g = 0.3$--0.35 and $-24^\circ \le \theta_P
\le -16^\circ$. Also, while the $\pi$ and $\eta$ production rates are
scale dependent, this dependence is very small and virtually
cancels out in the ratio, except for a mild dependence at smaller $p_T$
values, $p_T \lesssim 3$ GeV.

\begin{figure}[bhp]
\begin{center}
\includegraphics[trim = 0 20 250 280, width=0.55\textwidth,clip]{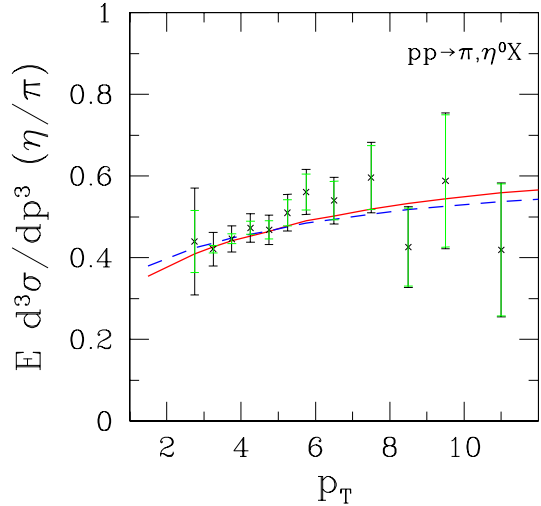}
\end{center}
\caption{The $\eta/\pi$ production ratios in $p\,p$ collisions as a
function of $p_T$ for $(f_g,\theta) = (0.3, -20^\circ)$ (solid line) and
$(0.35, -16.77^\circ)$ (dashed line) in comparison with
the RHIC/PHENIX data \cite{phenixeta} at $\sqrt{s} = 200$ GeV.}
\label{fig:etapi}
\end{figure}

The quality of fits to the $\eta/\pi$ ratio for the typical choices:
$(f_g,\theta_P) = (0.3, -20^\circ)$ and $(0.35, -16.77^\circ)$, are
shown in Fig.~\ref{fig:etapi} along with the RHIC/PHENIX data where both
statistical and total errors are shown.

\subsection{Fits to the 3-jet data in $e^+\,e^-$ collisions}

Fits to the individual quark fragmentation functions are of the same
quality as for the $\pi, K$ production discussed earlier (this data is
just the sum of the $\pi$ and $K$ rates). Hence we do not discuss them
further and proceed directly to gluon jet fragmentation.

With $f_g = 0.3$ as determined from fits to the $\eta/\pi$ ratio in
$p\,p$ collisions, it is now possible to compute the charged hadron
production rate from gluon fragmentation in $e^+\,e^- \to q \overline{q}
g$. We have, from Eq.~\ref{eq:gjet},
\begin{eqnarray} \nonumber
\left. \by{1}{N} \by{\d N^{\rm ch}}{\d x} \right\vert_g & = & 
\sum_h D_g^h~ \\
 & = & 2.9\hbox{--}3.2 D_g^\pi~, \nonumber
\end{eqnarray}
where we have assumed that $f_g^K = f_g^\eta = 0.3$--0.35, so that the
gluon fragmentation into $\eta$ and $K$ are roughly suppressed by the
same factor. If $\pi$ and $K$ production from gluon jets is separately
measured, $f_g$ (and its $x$-dependence, if any) can be easily
determined. Such data would be very useful in constraining further the
small-$p_T$ behaviour of the $\eta/\pi$ ratio.

\begin{figure}[htp]
\begin{center}
\includegraphics[trim=0 20 260 280,width=0.6\textwidth,clip]{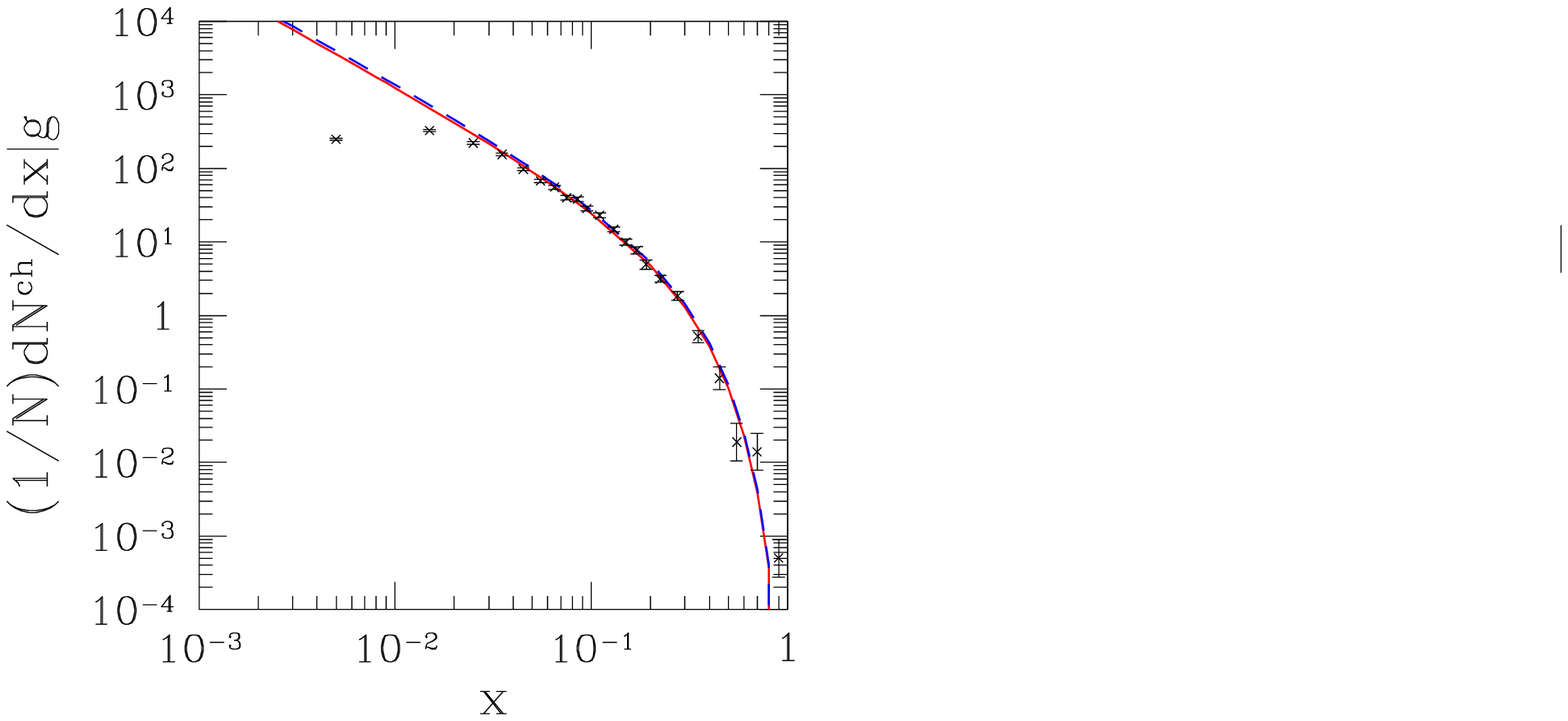}
\end{center}
\caption{The normalised charged hadron production rate from gluon jets
in $e^+\, e^- \to q \overline{q} g$ at a mean $Q^2 = 80^2$ GeV$^2$ at
LEP/OPAL \cite{OPAL} along with the theoretical prediction
$\sum_h D_g^h \sim 2.9$--$3.2 D_g^\pi$ where $h$ runs over all charged
hadrons and $K_S$. The lower (upper) curve corresponds to the choice of
2.9 (3.2), although the separation between them is not well-marked.}
\label{fig:opal_gl}
\end{figure}

The resulting fit to the OPAL data \cite{OPAL} is shown in
Fig.~\ref{fig:opal_gl}. Notice that the $Q^2$ scales of the $e^+\,e^-$
and $p\,p$ processes are very different: the former with a value $Q^2
\sim 80^2$ GeV$^2$ (slightly smaller than the $Z$-pole value for 2-jet
production), and the latter with $Q^2 = p_T^2 = 2^2$--14$^2$ GeV$^2$. Due
to the pole in the $P_{gg}$ splitting function, the gluon fragmentation
function is very different (in fact, at both large and small $z_h$
values) in the two cases and gives a strong consistency check on the
fits to the gluon fragmentation function.

\section{Discussion and Conclusions}

Quark and gluon fragmentation functions of $\pi$, $K$, $\eta$, and $\eta'$
mesons have been determined through a study of hadro-production rates in
both $e^+\,e^-$ and $p\,p$ collisions. Detailed information on the $\eta$
fragmentation functions has become important since the variation of the
$\eta/\pi$ production ratios in $p\,p$, $p\,A$ and $A\,A$ collisions is
a possible signal of quark gluon plasma (QGP).

All calculations have been performed at leading order (LO) in QCD.
Inclusion of NLO terms should not affect the quality of the fits,
especially that of the gluons, since they have been mostly determined
from ratios (normalised charged hadro-production in $e^+\,e^- \to$
3-jets and $\eta/\pi$ production ratio in $p\,p$ collisions). Of course,
numerical details, such as the fits to the quark and gluon fragmentation
functions at the starting scale $Q_0^2 = 2$ GeV$^2$, will change.
The fits therefore are mainly driven, and limited by, {\em the model
assumptions}. 

A simple SU(3) model for octet meson production ($\pi, K$), has been
extended to include $\eta$ and $\eta'$ production. Due to SU(3)
symmetry, the quark fragmentation functions of all mesons are related to
a common valence $V(x, Q^2)$ and sea quark $\gamma(x, Q^2)$
fragmentation function. SU(3) breaking due to the heavier strange quark
is included through a constant parameter $\lambda$. Inclusion of
singlet--octet mixing causes SU(3) breaking in the sea sector as well;
this is parametrised by a nonet mixing angle $\theta_P$ that is known
to be small ($-24^\circ \le \theta_P \le -10^\circ$), and a few additional
constants, $\lambda_8$, $\lambda_1$, and $f_d$. 

While $\lambda$, $\lambda_8$ and $\lambda_1$ are fixed from model
considerations, the data, especially on the $K/\pi$ ratio in $e^+\, e^-$
collisions, also yields the same value of $\lambda$ as that predicted by
the model. The parameter $f_d$ that occurs in the singlet fragmentation
functions is fitted from $\eta$ and $\eta'$ data in $e^+\,e^-$ collisions
on the $Z$-pole.

Finally, suppression of gluon fragmentation in $K, \eta$ relative to
$\pi$ is parametrised by $f_g(x)$. In the absence of more detailed data
(for instance, on individual $\pi$, $K$ production from gluon
fragmentation in 3-jet processes at LEP), $f_g$ is taken to be constant
and fitted from the large-$p_T$ ($2 \le p_T \le 11$ GeV) $\eta/\pi$
ratio in $p\,p$ collisions at $\sqrt{s} = 200$ GeV.

This gluon parametrisation, with $f_g \sim 0.3$--0.35, is completely
consistent with 3-jet data in $e^+\,e^-$ collisions at a very different
scale $Q^2 \sim 80^2$ GeV$^2$, where the gluon fragmentation is
explicitly studied.

A comprehensive set of gluon and quark fragmentation functions (for all
favlours) is therefore available for the set of nonet pseudo-scalar
mesons, $\pi^{\pm,0}$, $K^\pm$, $K^0$, $\overline{K}^0$, $\eta$, and
$\eta'$, starting with a simple model with reasonable assumptions, that
is consistent with hadro-production data in both $e^+\,e^-$ and $p\,p$
collisions. The focus of this paper was the extension of the earlier
fits \cite{IMR} in $\pi$ and $K$ to the case of $\eta$ and $\eta'$ mesons.

\paragraph{Acknowledgement}: We thank H.S. Mani and M.V.N. Murthy for
many detailed discussions.

\end{document}